\documentclass[graybox]{svmult}

\usepackage{mathptmx}       % selects Times Roman as basic font
\usepackage{helvet}         % selects Helvetica as sans-serif font
\usepackage{courier}        % selects Courier as typewriter font
\usepackage{type1cm}        % activate if the above 3 fonts are
\usepackage{makeidx}         % allows index generation
\usepackage{graphicx}        % standard LaTeX graphics tool
\usepackage{multicol}        % used for the two-column index
\usepackage[bottom]{footmisc}% places footnotes at page bottom

\makeindex             % used for the subject index
\begin{document}

\title*{Theory Overview of Testing Fundamental Symmetries}
\author{Nick E. Mavromatos}
\institute{Nick E. Mavromatos
\at King's College London, Department of Physics, Strand, London WC2R 2LS, UK. \\Currently also at: CERN, Physics Department, Theory Division, CH-1211 Geneva 23, Switzerland. \email{nikolaos.mavromatos@kcl.ac.uk}}
%\and Name of Second Author \at Name, Address of Institute \email{name@email.address}}
\maketitle
\vspace{-3cm}
\abstract{I review first some theoretical motivations for violation of Lorentz and/or CPT Invariance. 
Although the latter symmetries may be violated in a quantum gravity setting, nevertheless there are situations  in which
these violations are due to a given classical background geometry that may characterised early epochs of our Universe, and in fact be responsible for the observed dominance of matter over antimatter in the Universe. In this way I estimate some of the coefficients of the Standard Model Extension (SME), which is a framework for a field theoretic study of such a breakdown of fundamental symmetries. Then I describe briefly some tests of these symmetries, giving emphasis in low-energy antiproton physics and electric dipole moment measurements, of interest to this conference.  I also mention the r\^ole of entangled states of neutral mesons in providing independent measurements of T(ime reversal) and CP Violation, thus providing independent tests of CPT symmetry, as well as novel (``smoking-gun'' type) tests of decoherence-induced CPT violation, which may characterise some models of quantum gravity.}

\section{Theoretical motivations for Lorentz and CPT Violation}
\label{sec:theory}

The theory of Quantum Gravity is still elusive and far from any experimental verification.
Nevertheless, in the last decade there have been significant improvements in the precision of terrestrial and astrophysical instrumentation, which resulted in stringent bounds being placed on several models of quantum gravity available in the literature so far. Most of these models predict a breakdown of fundamental symmetries, such as (local) Lorentz invariance and CPT symmetry~\cite{review}. 

On the other hand, it is also possible that our Universe, at early epochs, was not characterised by the Robertson-Walker geometry, but by other, non-trivial background space times which may violated Lorentz and CPT symmetries in such a way that some remnants of such violations remained today, albeit small. In the latter case, the relics of such violations today may not be necessarily suppressed by the Planck scale, thus having a greater chance of being observed in the foreseeable future. In the first part of the talk I will discuss such a case, with the hope of better motivating the experimental searches for Lorentz and CPT violation. 

\emph{Lorentz \& CPT Violating Geometries in the Early Universe}: The simplest example of such a geometry, is provided~\cite{emsH} by an extension of the gravitational sector of the standard model to include a Kalb-Ramond (KR) antisymmetric tensor field of spin 1, $B_{\mu\nu}=-B_{\nu\mu}$. Such extensions, arise naturally in the low-energy limit of string theories, where the field $B_{\mu\nu}$ is the spin-one member of the massless gravitational multiplet of the string, the other two being the spin-zero (dilaton) and spin-two (graviton) fields~\cite{polchinski}. There is an abelian gauge symmetry characterising string theories, which persists in the effective low-energy local field-theory, namely $B_{\mu\nu} \to B_{\mu\nu} + \partial_{\left[\mu \, \right.}\, \theta_{\left. \nu \right]}$, where $\left[\mu, \nu\right]$ denotes proper antisymmetrization of the indices.  This invariance implies that the KR tensor appears in the effective field theory only through its field strength $H_{\mu\nu\rho} = \partial_{\left[\mu \, \right.} \, B_{\left. \nu \rho\right]}$.
In four dimensional space-times, obtained after appropriate compactification of the extra dimensions, the $H$ tensor can be expressed in terms of a pseudo scalar field $b(x)$ (KR ``gravitational axion'') 
$H_{\mu\nu\rho} = \epsilon_{\mu\nu\rho\sigma} \, \partial^\sigma b (x)~, \, \mu, \nu, \dots = 0, \dots 3,$
where $ \epsilon_{\mu\nu\rho\sigma} $ is the totally antisymmetric symbol in four space-time dimensions. 
It can be shown that the field strength $H_{\mu\nu\rho}$ plays a r\^ole analogous to a totally antisymmetric torsion, 
entering through a generalised Christoffel connection $\overline{\Gamma}_{\mu\nu}^\rho \ne \overline{\Gamma}_{\nu\mu}^\rho $:
$\overline{\Gamma}^\lambda_{\,\,\,\mu\nu} = \Gamma^\lambda_{\,\,\,\mu\nu} +  H^\lambda_{\mu\nu} \equiv
\Gamma^\lambda_{\,\,\,\mu\nu} + T^\lambda_{\,\,\,\mu\nu},$
where $\Gamma^\lambda_{\,\,\,\,\mu\nu} = \Gamma^\lambda_{\,\,\,\,\nu\mu} $ is the torsion-free Einstein-metric connection, and $T^\lambda_{\,\,\, \mu\nu} = - T^\lambda_{\,\,\,\nu\mu}$ is the torsion. 
Using general covariance,
the four-dimensional Lagrangian ${\mathcal L}_f $ for a spin-1/2 Dirac fermion $\psi$ in this background  can be cast in the form:
\begin{equation}\label{Bvector}
\mathcal{L}=\sqrt{-g}\,\overline{\psi}\left(i\gamma^{a}\partial_{a}-m+\gamma^{a}\gamma^{5}B_{a}\right)\psi~,\quad B^{d}
=\epsilon^{abcd}e_{b\lambda}\left(\partial_{a}e_{\,\, c}^{\lambda}+\overline{\Gamma}_{\nu\mu}^{\lambda}\, e_{\,\, c}^{\nu}\, e_{\,\, a}^{\mu}\right)~,
\end{equation}
above we have kept the space time curved since we are interested in situations pertaining to the early Universe where space-time may be non flat. In fact one may consider the Robertson-Walker background in the presence of the KR torsion. 

We observe from (\ref{Bvector}) that the non trivial 
background has the effect of
inducing an `axial' background field $B_{a}$ that has two contributions: one comes from the space-time curvature itself, 
and is proportional to derivatives of the villeins. This part vanishes for Robertson-Walker space times, but 
is known to be non-trivial
in certain anisotropic space-time geometries, such as Bianchi-type
cosmologies or in regions of space-time near rotating (Kerr) primordial black 
holes~\cite{Debnath:2005wk,Mukhopadhyay:2005gb,Sinha:2007uh}. 
The other contribution is proportional to the torsion, $ e_{b\lambda} \, \overline{\Gamma}_{\nu\mu}^{\lambda}\, e_{\,\, c}^{\nu}\, e_{\,\, a}^{\mu} $, and is non trivial in the presence of the KR field even in flat or Robertson-Walker space times. 
In fact, for a Robertson-Walker cosmological background, it can be shown that an exact solution in string theory~\cite{aben} necessitates a KR ``axion'' background  which is linear in the cosmic time $t$, thereby implying a \emph{constant}  $B^0$ background vector field in the comoving observer's frame of the expanding Universe. 

The magnitude of the vector $B^0$ depends on several parameters of the microscopic theory, among which the string scale. Essentially, it is a phenomenologically parameter, which may be substantially larger than the Planck scale. 
For constant $B^d$ vector, of which the constant $B^0$ vector in the case of $H$-torsion is a special case, the effective fermionic lagrangian (\ref{Bvector}) falls into the category of the simplest extensions of the standard model (SME)~\cite{kostel}, with a Lorentz and CPT violating fermion-axial vector interaction, the so-called $b_\mu$-vector in the SME formulation :
\begin{eqnarray}\label{sme}
{\mathcal L}_{SME} &\ni & \frac{1}{2} i \overline{\psi} \, \Gamma^\nu \, \partial_\nu \psi - {\overline \psi}\, {\mathcal M} \, \psi ~, \nonumber \\
{\mathcal M} &\equiv& m + a_\mu \gamma^\mu + b_\mu \, \gamma^5 \gamma^\mu + \frac{1}{2} H^{\mu\nu} \sigma_{\mu\nu}~, 
\quad \sigma_{\mu\nu} = \frac{1}{4} [ \gamma_\mu, \, \gamma_\nu ]~, 
\nonumber \\
\Gamma^\nu &\equiv & \gamma^\nu + c^{\mu\nu}\, \gamma_\mu + d^{\mu\nu}\, \gamma_5 \, \gamma_\mu + e^\nu + i f^\nu \, \gamma_5 + 
 \frac{1}{2} g^{\lambda \mu \nu} \, \sigma_{\lambda \mu}, ~.
\end{eqnarray} 
For our purposes we note that the terms proportional to $a_\mu$ and $b_\mu$ violate both Lorentz and CPT symmetries, unlike the $c_{\mu\nu}, d_{\mu\nu} $ and $H_{\mu\nu}$ terms that violate only Lorentz symmetries. Analogous observations can be made for the various terms inside the $\Gamma^\mu$ structure of (\ref{sme}). In the next section we shall discuss some phenomenological consequences of some of the above coefficients, especially in the context of neutral mesons, forbidden transitions in (anti)hydrogen molecules or atomic dipole moment measurements. 

At this point we would like to discuss in a bit more detail the important cosmological consequences of the $H$-torsion-induced $b^0$ nackground vector, which would also allow for some estimates of its value in the early Universe geometries. 
To this end, we remark, following the analysis in \cite{emsH}, that the presence of a constant (in a given frame) $B^0$ in (\ref{Bvector}) affects the dispersion relations of fermions, which are different from those of antifermions by terms of order $B^0$, thereby leading to induced CPT
violation. This, in turn, leads~\cite{emsH} to different populations between fermions and antifermions in the early universe, already in thermal equilibrium. In the model of \cite{emsH}, it is the massive right-handed (Majorana) (MRH) neutrinos that occur in extensions of the standard model, which oscillate between themselves and the corresponding antineutrinos, as a result of the presence of the (CPT Violating) $B^0$ background~\cite{Sinha:2007uh}, and which provide the chemical equilibrium (CP Violating) processes.
The MRH dispersion relations are: $E=\sqrt{(\vec{p})^{2}+m^{2}}+B_{0}~,\quad \overline{E} =\sqrt{(\vec{p})^{2}+m^{2}}-B_{0},$ where $E$ denotes the energy and the barred quantities refer to antiparticles. The oscillations freeze out at a given temperature $T_d$, which implies that below such temperatures, there will be a (frozen) difference of the populations of MRH neutrinos/antineutrinos, proportional to $B^0$, which in turn will lead, through appropriate decays, to Lepton number violation
\begin{equation}
\Delta L (T < T_d ) \sim \frac{B^0}{T_d}~.
\end{equation}
This is eventually communicated to the baryon sector via B-L conserving sphaleron processes, to produce the observed baryon asymmetry of ${\mathcal O}(10^{-10})$, provided $B^0 \sim 0.1$~GeV, $T_d \sim 10^9 $ GeV. These values depend on the details of the model, which will not be given here~\cite{emsH}. 
At $T_d$ the Universe is assumed to undergo a phase transition to $B^0 = 0$ or to a much smaller value $B^0$, that may survive until the present era. In the latter case it can be constrained by the phenomenology of the SME, which we shall consider in the next section. In particular, if a small $B^0$ survives today in the Robertson-Walker (or Cosmic-Microwave-Background (CMB)) frame, then in any boosted frame, with a velocity $\gamma \, {\vec v}$ with respect to that frame, one would also obtain a spatial background vector $\gamma \, {\vec v} \, B^0$ which can be constrained by precision atomic spectroscopy in both atoms and antiatoms, or comparison between the two, as we shall discuss in the next section.

\emph{Quantum-Gravity-Induced Lorentz and/or CPT Violation:} Before doing so, I would like to make the important remark that SME low-energy models may arise also as a result of quantum fluctuations of space time at microscopic (Planck) scales.
The sensitivity of various experiments to the so-called Physics at the ``Planck scale'', that is the energy scale at which quantum gravity phenomena are expected to set in, is highly model dependent. For instance, in the modern version of string theory~\cite{polchinski}, which is one of the most popular and thoroughly worked out theoretical frameworks of Quantum Gravity, the quantum gravity scale may not be the familiar one of Planck energy  $M_P = 1.2 \times 10^{19}$ GeV. The string mass scale, $M_s$, where quantum gravity phenomena take place in the higher-than-four-dimensional space times of string theory, is essentially unconstrained theoretically at present, and may be as low as a few TeV. This prompted the excitement for searching for effects of extra dimensions at the Large Hadron Collider (LHC), which has been recently launched at CERN. However, even if the scale of quantum gravity is as high as $M_P \sim 10^{19} \, {\rm GeV}$, nevertheless there may be predictions that affect the physics at lower scales, especially in models in which the quantum gravitational interactions behave as a `medium' (environment) in which ordinary matter propagates. The medium idea for quantum gravity is primarily due to J.A. Wheeler~\cite{wheeler}, who visualized Space-time at length scales near the Planck length $\ell_P \sim 10^{-35}$~m as having a ``foamy'' structure, that is containing singular quantum fluctuations of the metric field, with non trivial topologies of microscopic size, such as virtual black holes \emph{etc}.
These (defect) space-time structures may induce the breakdown of fundamental symmetries, in particular local Lorentz invariance and/or CPT symmetry.

The reader should bear in mind that in a model we may have Lorentz-invariance violating effects, but without any CPT Violation in the Hamiltonian~\cite{lehnert}. An example is provided by the so-called non-commutative field theory models. In some of them, one can argue~\cite{carrol} that their low-energy continuum space-time description corresponds to effective field theories of the form encountered in the so-called standard model extension of Kostelecky and collaborators~\cite{lehnert} but with only Lorentz violating higher-dimensional operators, while CPT appears unbroken by the effective lagrangian. On the other hand, if CPT is violated, then both Lorentz- and CPT -symmetry violating effects are present. This seems to be a general consequence of
the axiomatic proof of CPT theorem in flat space time models~\cite{greenberg}, which requires
Lorentz-covariant off-shell correlation functions in a relativistic field theory setting. Note however the counterexample of ref. \cite{chaichian}, in which Lorentz invariant CPT Violation has been demonstrated in non-local field theory models where a transfer matrix is not well defined (which was one of the main assumptions of the theorem of \cite{greenberg}). 

In this talk, I will concentrate mainly on the breakdown of the CPT symmetry. 
As we have seen above, the latter may be induced either by a (Lorentz violating) background space-time describing the Early universe, or by quantum effects of the vacuum including gravitational fluctuations. 
In the latter case, there are two ways by which CPT breakdown is encountered in a quantum gravity model. The first is through the non commutativity of a well-defined quantum mechanical CPT operator (which generates the CPT transformations) with the Hamiltonian of the system under consideration. This is the breakdown of CPT symmetry dealt with in standard Lorentz-violating Extensions of the Standard Model (SME), mentioned above~\cite{lehnert,greenberg,carrol}. 
In the second way of CPT breaking, the CPT operator is \emph{ill-defined} as a quantum mechanical operator, but in a \emph{perturbative sense} to be described below. This ill-definition is a consequence of the foamy structure~\cite{wheeler} of space time, whereby the quantum metric fluctuations at Planck scales induce \emph{quantum decoherence} of matter propagating in such backgrounds.
For such cases, the particle field theoretic system is simply an \emph{open quantum mechanical system} interacting with the ``environment'' (or ``\emph{medium}'') of quantum gravity. 

The feature of an ill-defined CPT operator in such cases is of a more fundamental nature than the mere non commutativity of this operator with the local effective Hamiltonian of the matter system in Lorentz-symmetry violating SME models. In the cases of quantum-gravity induced decoherence the very concept of a local effective Lagrangian may itself break down. R.~Wald~\cite{wald} has elegantly argued, based on elementary quantum mechanical analysis of open systems,
that the CPT operator cannot exist as a well-defined quantum mechanical operator for systems which exhibit quantum decoherence, that is they are characterised by an evolution of initially pure quantum mechanical states to mixed ones, as the time elapses. This was interpreted as a microscopic time arrow in quantum gravitational media, which induce such decoherence, that is a fundamental T violation unrelated to CP properties. We term this phenomenon  ``\emph{intrinsic CPT violation}''. 
As a result of the weak nature of quantum gravitational interactions, the ill-definition of the
CPT operator is \emph{perturbative} in the sense that the \emph{anti-particle state} still \emph{exists}, but its properties, as compared to the corresponding particle state, which under normal circumstances would be connected by the action of this operator, are modified. The modifications can be perceived~\cite{sarkar} as a result of the dressing of the (anti-)particle states by perturbative interactions expressing the effects of the medium. In such an approach, the Lorentz symmetry aspect is disentangled from the CPT operator ill-defined nature, in the sense that
Lorentz invariance might not be necessarily violated in such systems (we note that Lorentz-invariant decoherence is known to exist, in the sense of decohered systems with modified Lorentz symmetries, though, to take proper account of the open-system character~\cite{millburn}). 

An interesting question, of experimental interest, concerns the possibility of the observer to prepare decoherent-free subspaces in such quantum-gravity entangled systems. If such a possibility could be realized, then one would have a ``weak form of CPT invariance'' characterising
the system~\cite{wald}, in the sense that
the ill-defined nature of the fundamental CPT operator would not show up in any physical quantities measured in Nature, in particular scattering amplitudes.
Although, theoretically, such a possibility is still not understood, nevertheless the question as to whether there are decoherence-free subspaces in quantum-gravity foam situations can be answered experimentally, at least in principle.
It is also among the points of this review to tackle this issue by discussing the effects of  this ``intrinsic CPT violation'' in entangled states of mesons in meson factories. As argued in \cite{bmp}, the perturbatively ill-defined nature of  the CPT operator implies modified Einstein-Podolsky-Rosen (EPR) correlations among the entangled states in meson factories, which are uniquely associated with this effect and can be disentangled experimentally from conventional background effects. We termed this effect $\omega$-effect. My point is that, although in general there seems to be no single figure of merit for CPT Violation, as this is a highly model-dependent issue, nevertheless, these EPR correlation modifications, if true, may constitute ``smoking-gun'' evidence for this particular type of CPT violation and decoherence in Quantum Gravity.

The structure of the remainder of the talk is as follows: in section \ref{sec:2}, I describe some selected tests of Lorentz and CPT Violating extensions of the standard model, using antimatter factories or atomic dipole moments, of interest to this conference. Then I proceed, in section \ref{sec:3}, to a discussion on the r\^ole of entangled meson states as accurate probes of discrete symmetries, CP, T and CPT, giving emphasis on novel tests for Time Reversal (T) Violation measurements, independent of CP and CPT, and the novel $\omega$-effect associated with a potential quantum-gravity-decoherence-induced CPT Violation. Conclusions and outlook are presented in section \ref{sec:concl}.

\section{SME Tests using antimatter factories and atomic dipole-moment measurements \label{sec:2}}

Currently there exists an exhaustive literature for Standard Model Extensions (SME) in the fermion sector~\cite{lehnert,smeanti} (\ref{sme}), 
in the sense of writing down the most general Lorentz- (LV) and/or CPT-Violating (CPTV) terms which can produce interesting effects in delicate atomic physics precision experiments or antimatter factories, and from the non-observation of the respective terms one can place stringent upper limits to  the various coefficients. There is a plethora of tests, ranging from atomic spectroscopy in both matter and antimatter systems and atomic dipole moment measurements to neutrino oscillations, by which the SME coefficients can be bounded. In addition to such fermion-only theories, one has considered LV-and/or CPTV Extensions of the Standard Model in the gauge, scalar and fermion sectors, where higher-dimension (five) field operators that violate Lorentz symmetry and/or CPT have been classified fully~\cite{lvqed}.  The main assumption behind the form of such operators is that an unknown physics at high energy scales could lead to a spontaneous breaking of Lorentz invariance by giving an expectation value to certain tensorial fields, which are not in the Standard Model (SM) spectrum~\cite{lehnert}. The interaction of these fields with operators composed from the SM fields, which are fully
Lorentz-symmetric before the spontaneous breaking, 
will manifest itself as effective LV terms, which below the scale of the LV condensation would have the schematic form: 
\begin{equation}\label{LVqed}
O^{\rm. SM}_{\mu\, \nu \, \dots}  \, C^{\mu\, \nu \, \dots } \rightarrow O^{\rm SM} \, \langle C^{\mu\, \nu \, \dots } \rangle~,
\end{equation}
where $C^{\mu\nu \dots }$ is an external field that undergoes condensation and $O^{\rm SM}$ is a SM field operator
that transforms properly under the Lorentz group.
The classification of \cite{lvqed} requires that the independent dimension-5 operators must be gauge invariant, Lorentz invariant after contraction with the background tensors $\langle C^{\mu_1 \mu_2 \dots} \rangle $, not reducible to total derivatives or to lower-dimension operators by the use of equations of motion, and they should couple to an irreducible background tensor. 

Several experiments, of diverse origin, can be used in order to impose stringent constraints on the relevant SME coefficients, that range from searches for forbidden atomic transitions in precision experiments and studies of low-energy antiprotonic atoms and antimatter factories, to high-energy cosmic rays, nuclear spin precession and atomic and nuclear Electric Dipole Moments (EDM) measurements, as well as data on neutrino oscillations. 

In this section, we shall review briefly the current bounds on some SME coefficients coming from antimatter atomic transition
spectroscopy and EDM. 
We commence with SME tests in antiprotonic atoms~\cite{smeanti}, in particular antihydrogen (${\overline {\rm H}}$) of great interest in this conference. Motivated by the theoretical microscopic models  of section \ref{sec:theory}, I shall restrict myself for the purposes of this talk to constraining the $b_\mu$ coefficients of the SME (\ref{sme}) using spectroscopy, in particular looking for forbidden transitions, \emph{e.g}. $1s \to 2s$ . Within H spectrocopic measurements, the presence of a $b_\mu$ coefficient in the SME (\ref{sme}) leads to the 
relevant transition of the electron in the H atom. The sensitivity of the tests depend crucially whether the atoms are free or trapped in an external magnetic field. In the case of free H (and ${\overline {\rm H}}$), the frequency shift of the 1s-2s transition is a higher-loop quantum effect in the SME/Quantum-Electrodynamics (QED) lagrangian, and thus the effect is suppressed by the square of the fine structure constant, $\alpha^2$: $\delta _{1s-2s}\nu^{{\rm H}} \simeq -\alpha^2 \, b_3^e/8\pi $, \emph{i.e.} the pertinent sensitivity of such experiments would be about five orders of magnitude smaller compared to tests involving the corresponding transitions in trapped H and ${\overline {\rm H}}$. However, in the latter tests, the corresponding frequency  shifts are proportional to the difference 
$b_3^e - b_3^p$ of the third spatial component of $b_\mu$ between electrons (e) and protons(p) (in a frame where the direction of the external magnetic field is along the $z$ axis). 
In view of the universal character of $B^\mu$ vectors due to background space-time geometries 
discussed in section \ref{sec:theory}, for this model
the above difference would vanish. To cover ourselves against such cases, it is therefore imperative to either measure the sum of the coefficients $b_\mu^{e,p}$, 
or isolate them experimentally. The former can be achieved
by examining hyperfine structure transitions in atomic (anti)matter. Indeed, within 1s transitions of H or ${\overline {\rm H}}$,
one can determine the relevant energy shifts induced by $b_\mu$~\cite{smeanti}:
\begin{equation}\label{hypefine}
\Delta_{a\to b}^{H} \simeq (b_\mu^e + b_\mu^p)/\pi + \dots
\end{equation}
where the $\dots$ denote contributions from the rest of the SME coefficients (\ref{sme}), which are not written explicitly here. 
Hyperfine transitions within the 1S level of H can be measured with accuracies exceeding 1
mHz in masers. So transitions of this type in trapped H and ${\overline {\rm H}}$ are interesting
candidates for performing tests of Lorentz or CPT symmetry, although to achieve resolutions of 1 mHz in trapped antihydrogen does not seem feasible 
in the foreseeable future.

Another possibility would be to measure~\cite{smeanti} radio-frequency transitions between states within
the triplet of hyperfine levels in H and ${\overline {\rm H}}$, in particular the so called $d\rangle_1 \to |c\rangle_1$ transition at external magnetic fields of order B $\simeq$ 0.65 Tesla. The corresponding frequency shifts depend solely on $b_3^p$:
\begin{equation}
\Delta_{c \to d}^{\rm H} \simeq -b_3^p/\pi, \qquad \Delta_{c \to d}^{\overline {\rm H}} \simeq +b_3^p/\pi\end{equation}
where we took into account that under the action of CPT operation, which exchanges H and 
${\overline {\rm H}}$, the coefficient of the $b_3^p$ changes sign.  Thus, comparison of the above spectroscopic measurement 
between trapped H and ${\overline {\rm H}}$ would yield immediately a bound (or a value !) on $b_3^p$. If  a frequency resolution of 1 mHz could be attained (which at present is far from being plausible), then, one could obtained $|b_3^p | \le 10^{-27}$~GeV. Still such bounds are about four orders of magnitude smaller that the ones coming from masers. 
We also note that, although, clock-comparison experiments are able to resolve spectral lines to about 1 $\mu$Hz,
nevertheless, isolating  $b_i^p$ is very complicated due to the complex 
structure of the nuclei involved.  

The above experiments are sensitive only to spatial components of Lorentz-violating couplings. 
Sensitivity to
timelike couplings, $b^0$, would require appropriate boosts. On the other hand, in the context of the model (\ref{Bvector}) of section \ref{sec:theory}, such experiments can bound the combinations $\gamma {\vec v}_3 B^0$, where ${\vec v}$ is the current-era relative  velocity of us (as local observers) with respect to the CMB (or Friedmann-Robertson-Walker) frame. 

\begin{figure}
\centering
  \includegraphics[width=6.5cm]{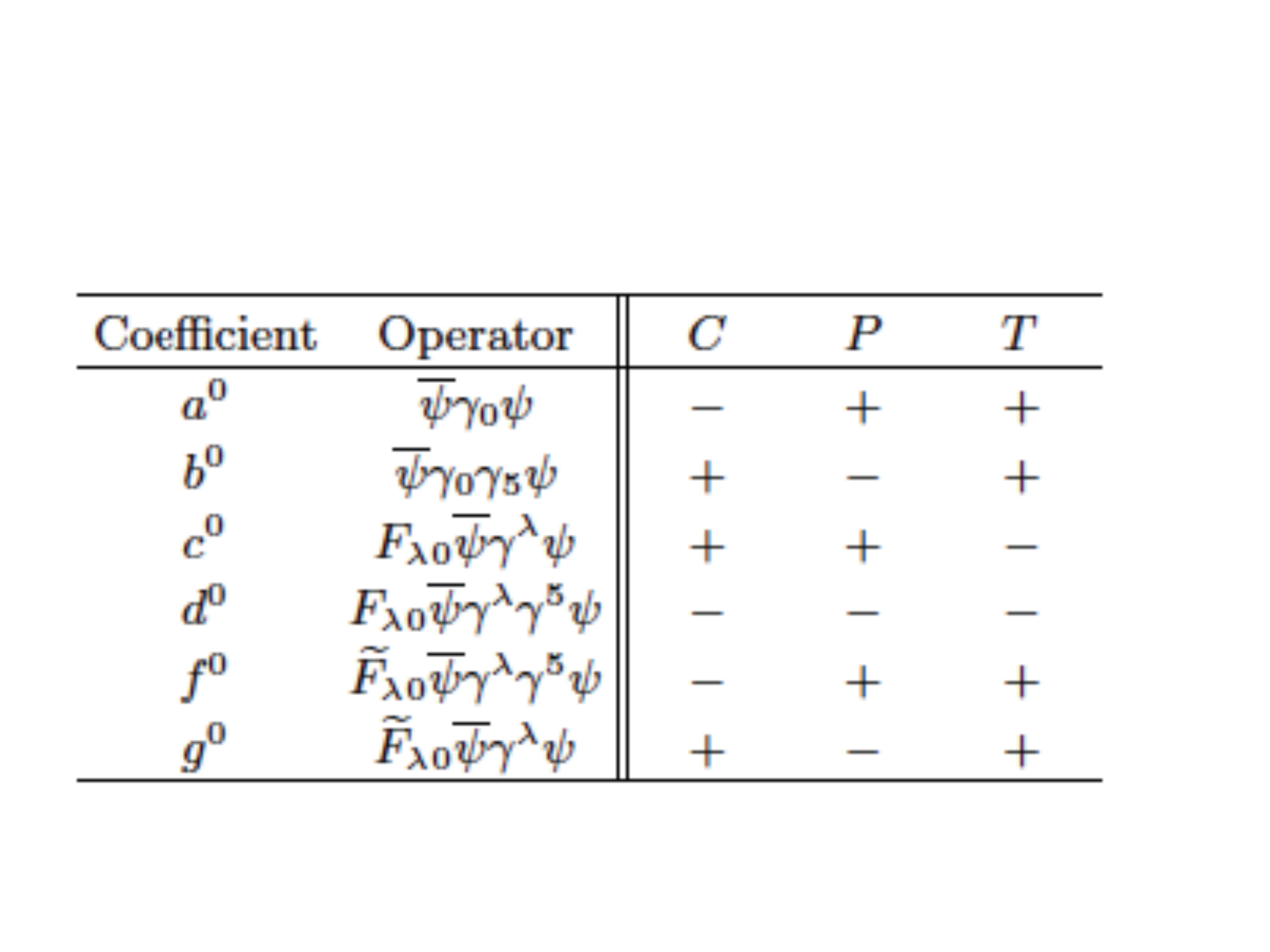}
\caption{Tranformation properties of the various terms in (\ref{qed5}) under the action of the discrete symmetries C, P and T. A + indicte even and a - odd. From ref. \cite{lvqed}.}
\label{fig:properties}
\end{figure}

Next we  describe the situation governing the constraints on the relevant dimension-5 terms of the SME lagrangian coming from EDM. These are generically found to be of order~\cite{lvqed}  $\le \, 10^{-25} \, e\, {\rm cm}$. 
The overall expression for the total EDM, due to the CP Violating conventional QED
terms and the CPT Violating terms due to the presence of an appropriate Lorentz-violating background vector $n^\mu$, is obtained from the effective Lagrangian
\begin{equation}\label{edm}
{\mathcal L}_{\rm EDM} = -i\frac{1}{2} d_{\rm CP} \overline{\psi} \, \sigma^{\mu\nu} \, F_{\mu\nu} (A) \psi + 
d_{\rm CPT} \overline{\psi} \, \gamma^\mu \, \gamma^5  \, F_{\mu\nu} (A) \, n^\nu\, \psi ~,
\end{equation}
where $F_{\mu\nu}$ is the Maxwell field strength. The currently null result on the neutron dipole moment imposes the constraint $d_{\rm CP} + d_{\rm CPT} = 0$. The lagrangian (\ref{edm}) should be completed with the $a^\mu$ and $b^\mu$ SME terms (\ref{sme}), as well as the appropriate dimension-5 operators from the QED sector of the SME~ \cite{lvqed}:
\begin{equation}\label{qed5}
{\mathcal L}_5 = \sum_{\rm fermion~species} \Big[c^\mu \, \overline{\psi} \gamma^\lambda F_{\lambda \mu} \psi + d^\mu\, \overline{\psi} \gamma^\lambda \, \gamma^5 \, F_{\lambda \mu} \psi  + g^\mu \overline{\psi} \gamma^\lambda {\tilde F}_{\lambda \mu} \psi +
 f^\mu \, \overline{\psi} \gamma^\lambda \, \gamma^5 {\tilde F}_{\lambda \mu} \psi \Big]~,
\end{equation}
where ${\tilde F}_{\mu\nu}$ is the dual of the Maxwell tensor. The various terms in (\ref{qed5}) have different transformation properties under the action of the discrete symmetries C, P and T, which, together with the corresponding terms of (\ref{sme}), are indicated in figure fig.~\ref{fig:properties}, on the assumption that the vector backgrounds are time-like and invariant under C,P and T reflections~\cite{lvqed}.

Experimentally~\cite{romalis}, one can disentangle CP-odd from CPT-odd operators, because 
of different suppression scales. Specifically, the former require helicity flip and are thus represented by dimenion-six operators in the SME effective lagrangian, with suppression by the CP breaking scale of order $1/\Lambda^2_{\rm CP}$. Such operators imply spin precession in a magnetic field relative to the direction of 
$\textbf{B} \times \textbf{v} $. On the other hand, the CPT-odd operators are of dimension 5, as they do not require helicity flip, \emph{e.g.} in the quark sector such operators are of the form
${\overline q}_{R(L)} \gamma^\nu \, \gamma^5 \, F_{\nu\, \mu}\, q_{R(L)}$, and ${\overline q}_{L} \gamma^\nu \, \gamma^5 \, F^a_{\nu\, \mu}\, \tau^a \, q_{L}$, where $\tau^a$, $a=1,2,3$ are the SU(2) generators of the weak interaction standard model group, and $F_{\mu\nu}$ and $F_{\mu\nu}^a$ are the U(1) and SU(2) gauge field strengths respectively. These operators are suppressed linearly by the CPT-breaking scale, $1/\Lambda_{\rm CPT}$.

EDMs have been bounded with high precission in~\cite{romalis}: 

(i) neutrons, with the bound $d_n \, < \, 3 \times 10^{-26} \, {\rm e\, cm}~$, 

(ii) diamagnetic atoms (such as Hg, Xe, ...) : their EDMs are induced by the EDMs of the valence nucleons; for the case of mercury EDM, one has the (approximate) relation: $d_{\rm Hg} \simeq -5 \times 10^{-4} \big(d_n + 0.1 \, d_p \big) \sim - 5 \times 10^{-4} \, d_n $. The last approximate relation implies that a signal consistent with CPT violation would occur, if a non zero $d_n$, $d_{\rm Hg}$ were found.

(iii) paramagnetic atoms (such as Tl, Cs, ...): their EDM are extremely suppressed as a result of the absence of a CPT-odd electron EDM.

In general, theoretical estimates of dimension-three operators induced by multiloop CP violating corrections in the standard model, imply the following bounds of the SME coefficients in (\ref{sme})~\cite{romalis} 
\begin{equation}\label{cpodd}
a^\mu, b^\mu \sim \, d^\mu \, \Big( 10^{-20} - 10^{-18} \Big) \, {\rm GeV}^2 ~, 
\end{equation}
providing sensitivity to $d^\mu \le 10^{-12}$~GeV$^{-1}$ and thus $\Lambda_{\rm CPT} \sim \Big(10^{11} - 10^{12} \Big)$~GeV.

Higher Lorentz-violating background tensors, \emph{e.g.} terms in SME effective lagrangian of the form
${\mathcal D}_{\mu\nu\rho} {\overline e} \, \gamma^\rho \, \gamma^5 \, e \, F_{\mu\nu} $ can also be bounded experimentally with high accuracy, by looking~\cite{romalis} for corrections to the spin precession frequency of the form $ \big({\mathcal D}^{i [0\, k]} + {\mathcal D}^{k \, [0\, i ]} \big) \, E_i \, B_k$, which changes sign under the reversal of the electric field $E_i$. The relative signal changes during the day as a result of the change of the Laboratory orientation relative to the tensor background. 

\begin{figure}
\centering
  \includegraphics[width=5.5cm]{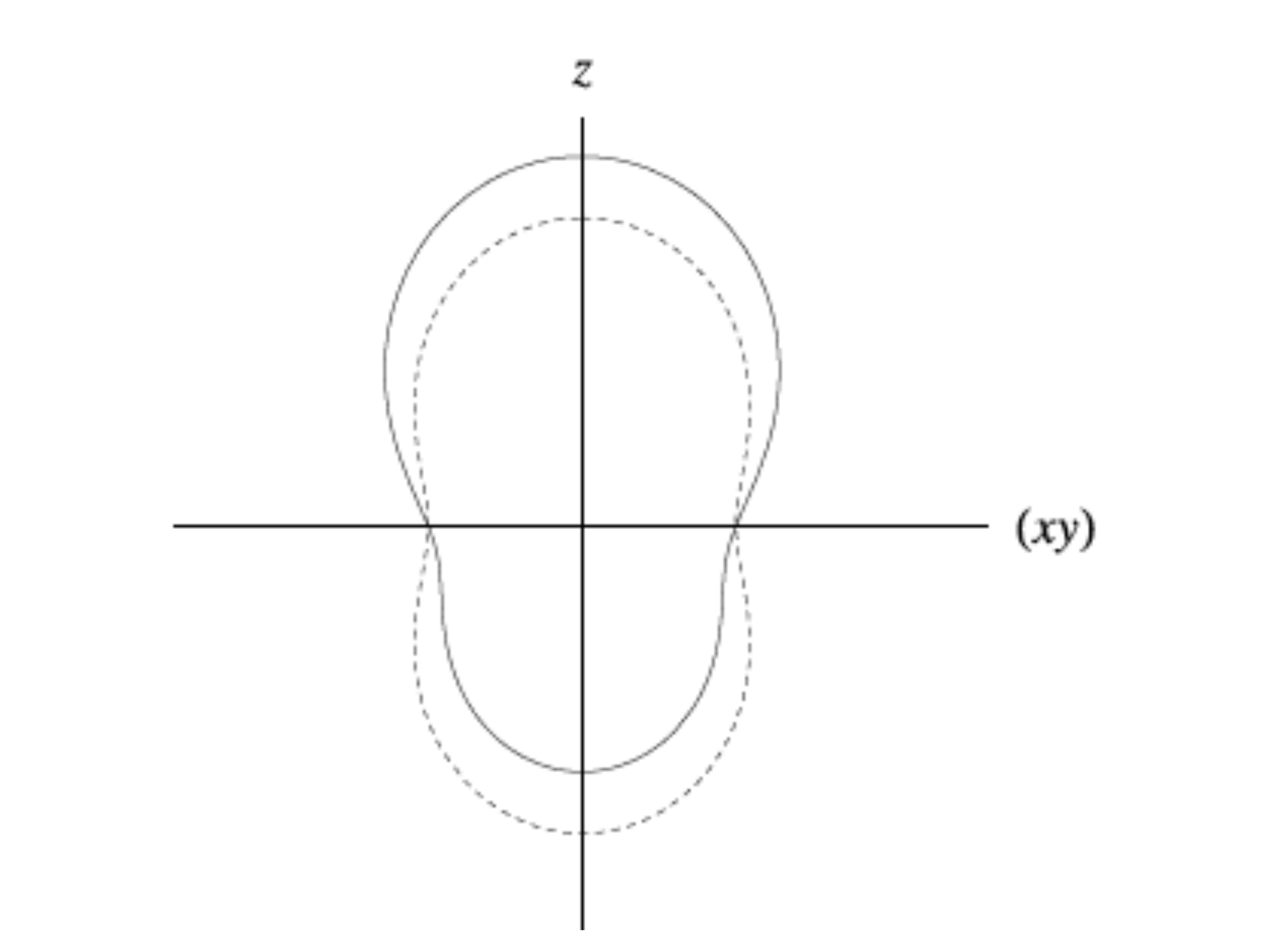}
\caption{Anglular distribution for spontaneous radiation for the atomic transition $2p_{1/2,1/2} \, \to \, 1s_{1/2, -1/2}$ in the presence of a CPT-odd SME axial background vector $b^\mu$. The dashed line indicates the standard electrodynamics $b^\mu=0$ case. From ref.~\cite{edmfoldy}.}
\label{fig:anapole}
\end{figure}

We close this section by mentioning the interesting suggestion of ref. \cite{edmfoldy} on further tests of CPT symmetry due to the CPT-odd axial vector background $b^\mu$, which has been of interest to us in section \ref{sec:theory}. According to this work, within the framework of Lorentz-violating extended electrodynamics, the Dirac equation for a
bound electron in an external electromagnetic field has been considered, assuming the interaction with
the background field $b_\mu$. A Foldy-Wouthysen  quasi-relativistic (1/c)-series expansion (truncated to order $1/c^2$)  has been applied to obtain an effective Hamiltonian for the hydrogen atom and through this the relativistic Dirac eigenstates in a spherically-symmetric potential  to second order in $b^0$. The $b^0$-induced CPT-odd corrections to the electromagnetic dipole moment operators of a bound electron have been calculated. Such corrections  contribute to the anapole moment of the
atomic orbital and may cause a specific asymmetry of the angular distribution of the radiation of a
hydrogen atom, in particular the $2p_{1/2,1/2} \, \to \, 1s_{1/2, -1/2}$ (\emph{cf.} fig.~\ref{fig:anapole}). 
The non-observation currently of such asymmetries leads to bounds of the magnitude of $|b^0|$: $|b^0| \, \le \, 2 \times 10^{-8} \, m_e \, c^2  \simeq  10^{-11} \, {\rm GeV}$, which are consistent with the general bounds for the SME coefficient $b_\mu$ for electrons~\cite{lehnert} $b^0 \le 0.02 \, {\rm eV}$ and $|{\textbf b}| \, \le 10^{-19} $~eV. 
Such tests may also be performed in man-made antihydorgen or other anti-atoms, with the aim of providing direct comparison of CPT properties and thus tests of CPT invariance. 

Finally we mention that, further tests of CPT invariance can be made by direct measurements of 
particle antiparticle mass and charge differences, which we are not going to discuss here. 
However, in the spirit of our cosmological model discussed in section \ref{sec:theory}, we do mention that, if 
 the observed matter/antimatter asymmetry were due to a mass difference between particle and antiparticles, then, one may make the reasonable assumption that baryogenesis could be due to mass differences between quarks and antiquarks~\cite{dolgov}. The latter nay depend linearly with temperature, $m_q(T) \sim g T$, as a consequence of known high-temperature properties of Quantum Chromodynamics (QCD).  Furthermore, it is reasonable (although not strictly necessary) to assume that the quantk-antiquark differences today are bound by the current bound on proton-antiproton mass difference, which is of order $ 7 \times 10^{-10}$ GeV, as provided in 2011 by the ASACUSA Collaboration~\cite{asacusa}. Scaling back in temperature such differences, up to the respective decoupling temperature of the quarks, lead to baryon asymmetries that are much smaller than the observed one~\cite{dolgov}. In this sense the model of \cite{emsH} can still survive, given that, even if a $B^0 < 0.02 $ eV is observed today, according to the current SME limits, the Universe may have undergone such a (or series of) phase transition at $T \sim 10^9 $ GeV towards a smaller (or zero) H-torsion background. This is an (crude) example of how one can use current SME bounds to fit early universe cosmologies.

\section{Testing Fundamental Symmetries in Entangled Meson Factories}
\label{sec:3}

In this section we move onto a discussion of CPT-Violating terms in the SME and beyond (namely, quantum gravity-decoherence induced situations that cannot be described as local effective field theories) in facilities involving entangled states of neutral mesons, such as neutral Kaon($\Phi$) factories~\cite{adidomenico}
or $B-{\overline B}$ meson factories~\cite{babar}.
 
We commence our discussion by briefly mentioning direct tests  of Time reversal invariance within the Lorentz invariant standard model theory, using entangled neutral mesons, independently of CP and CPT violation. These have been initially proposed in \cite{entangledt}, leading to the recent observation of \emph{direct} T violation by the Ba-Bar collaboration~\cite{babar}, through the exchange of initial and final states in transitions that can only be connected by a T -symmetry transformation.
For example,  the transition ${\overline B}^0 \, \to \, B^- $ for the second B to decay, at time $t_2$, once the first B (entangled with the second) has been tagged at time $t_1$, is identified by reconstructing events in the time-ordered final states $(\ell^+ \, X, J/\psi \, K_s^0)$. The rate of this transition is then compared to 
that of the $B^- \, \to \, {\overline B}^0$ transition, that exchanges initial and final states, which is identified by the reconstruction of the final states $(J/\psi \, K_L^0, \ell^- \, X )$. Any observed difference between these two rates, would thus indicate direct observation of T violation, independent of CP properties.
This would also imply an independent test of CPT symmetry within the standard Model. 
Similar tests of T violation in entangled Kaon $\Phi$ factories have also been suggested~\cite{kaont}, by identifying the appropriate reactions that exchange initial and final states.

However, if CPT is \emph{intrinsically} violated,
in the sense of being not well defined due to decoherence~\cite{wald} induced by quantum gravity~\cite{ehns}, the above-mentioned direct observation of T violation cannot constitute a test of decoherence-induced CPT breaking. This is because in such a case a distinct phenomenon, associated with the ill-defined nature of CPT operator, emerges, termed $\omega$-effect~\cite{bmp}. 

\begin{figure}
\centering
  \includegraphics[width=3.5cm]{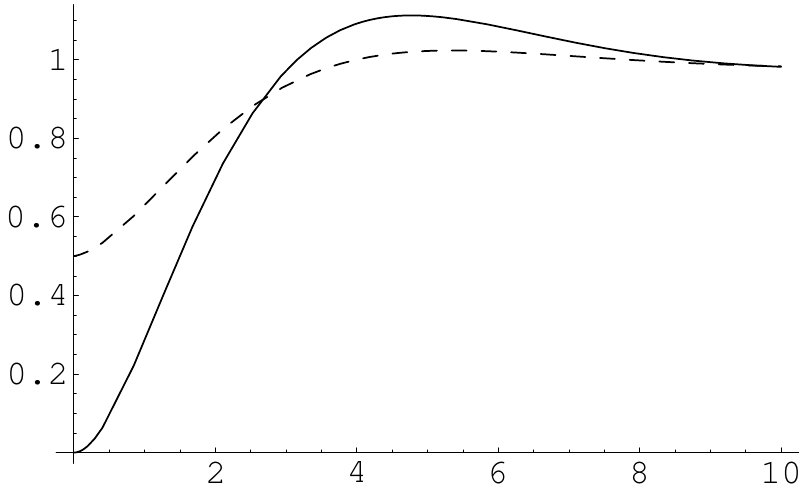}
\caption{A characteristic case of the intensity
$I(\Delta t)$, with $|\omega|=0$ (solid line)  vs  $I(\Delta t)$
(dashed line) with $|\omega|=|\eta_{+-}|$, $\Omega = \phi_{+-} -
0.16\pi$, for definiteness~\cite{bmp}.}
\label{intensomega}
\end{figure}

Let us concentrate for simplicity to the neutral Kaon system, where the effects as we shall see are dominant, although conceptually our analysis applies equally~\cite{bfactories} to entangled B-meson factories as well, such as those of \cite{babar}. In a quantum-gravity induced decohered situation, the Neutral mesons $K^0$ and ${\overline K}^0$ should \emph{no longer} be treated as \emph{identical particles}. As a
consequence~\cite{bmp}, the initial entangled state in $\Phi$
factories $|i>$, after the $\Phi$-meson decay, assumes the form:
{\scriptsize $ |i> = {\cal N} \bigg[ \left(|K_S({\vec
k}),K_L(-{\vec k})>
- |K_L({\vec k}),K_S(-{\vec k})> \right)\nonumber +  \omega \left(|K_S({\vec k}), K_S(-{\vec k})> - |K_L({\vec
k}),K_L(-{\vec k})> \right)  \bigg]$}, where $\omega = |\omega |e^{i\Omega}$ is a complex parameter,
parametrizing the intrinsic CPTV modifications of the EPR
correlations~\cite{bmp}. The $\omega$-parameter controls the amount of contamination of the
final C(odd) state by the ``wrong'' (C(even)) symmetry state.
The appropriate observable (\emph{c.f}. fig.~\ref{intensomega})
is the ``intensity'' $I(\Delta t)
= \int_{\Delta t \equiv |t_1 - t_2|}^\infty
|A(X,Y)|^2$, with $A(X,Y)$ the appropriate $\Phi$ decay
amplitude~\cite{bmp},
where one of the Kaon products decays to
the  final state $X$ at $t_1$ and the other to the final state $Y$
at time $t_2$ (with $t=0$ the moment of the $\Phi$ decay).
It must be noticed that in Kaon factories there is a particularly good channel,
the one with bi-pion states $\pi^+\pi^-$ as final decay products,
which \emph{enhances the sensitivity}
to the $\omega$-effect by three orders of magnitude. This is due to the fact that
the relevant terms~\cite{bmp} in the intensity $I(\Delta t)$ (c.f. fig.~\ref{intensomega})
contain the combination $\omega/|\eta_{+-}|$, where $\eta_{+-}$ is the relevant CP-violating
amplitude for the $\pi^+\pi^-$ states, which is of order $10^{-3}$.
The KLOE experiment has just released the first measurement of the
$\omega$ parameter~\cite{adidomenico}: $ {\rm Re}(\omega) =
\left( -2.5^{+3.1}_{-2.3}\right)\times 10^{-4}~$, ${\rm
Im}(\omega) = \left( -2.2^{+ 3.4}_{-3.1}\right)\times
10^{-4}$. At least an order of magnitude improvement is expected
for upgraded facilities such as KLOE-2 at (the upgraded)
DA$\Phi$NE-2~\cite{adidomenico}. This sensitivity is not far from
certain optimistic models of space time foam leading to
$\omega$-like effects~\cite{sarkar}.

In B-factories one can look for similar $\omega$-like effects.
Although in this case there is no particularly good channel to lead to enhancement of the sensitivity, as in the $\Phi$-factories, nevertheless one gains in statistics, and hence interesting limits may also be obtained~\cite{bfactories}. The presence of a quantum-gravity induced $\omega$-effect in B systems is associated with a theoretical limitation on flavour tagging, namely the fact that in the absence of such effects the knowledge that one of the two-mesons in a meson factory decays at a given time through a flavour-specific "channel'' determines unambiguously the flavour of the other meson at the same time. This is not true if intrinsic CPT Violation is present. One of the relevant observables~\cite{bfactories} is given by the CP-violating semi-leptonic decay charge asymmetry (in equal-sign dilepton channel), with the first decay $B \to X\ell^{\pm}$ being time-separated from the second decay
$B \to X'\ell^{\pm}$ by an interval $\Delta t$.
In the absence of $\omega$-effects, the intensity at equal decay times vanishes, $I_{\rm sl}(\ell^{\pm},\ell^{\pm},\Delta t=0) = 0$, whilst in the presence of a complex $\omega=|\omega|e^{i\Omega}$, $I_{\rm sl}(\ell^{\pm},\ell^{\pm},\Delta t=0) \sim |\omega |^2$. In such a case, the asymmetry observable  
exhibits a peak, whose position depends on $|\omega|$, while the shape of the curve itself depends on the phase $\Omega $~\cite{bfactories}. The analysis of \cite{bfactories}, using the above charge asymmetry method and comparing with currently available experimental data, leads to the following bounds:
$ -0.0084 \le {\rm Re}(\omega) \le 0.0100 $ at 95\% C.L.. Such tests for intrinsic CPT violation may be performed simultaneously with the above-mentioned observations of direct T violation, as they are completely independent. 

Before closing we would  also like to point out that an observation of the $\omega$-effect in both the $\Phi$ and B-factories could also provide an independent test of Lorentz symmetry properties of the intrinsic
CPT Violation, namely whether the effect respects Lorentz symmetry.
This is because, although the $\Phi$ particle in neutral Kaon factories is produced at rest,
the corresponding $\Upsilon$ state in B-factories is boosted, and hence there is a frame change between the two experiments. If the quantum gravity $\omega$-effect is Lorentz violating, as it may happen in
certain models~\cite{sarkar}, then a difference in the value of $\omega$ between the two
experiments should be expected.

Finally, since we mentioned Lorentz Violation, we also point out that bounds of the LV SME coefficients $a^\mu$ (\emph{cf}. eq.~(\ref{sme})) can be placed by measurements in the entangled Kaon $\Phi$ factories~\cite{adidomenico}. In particular by adopting the relevant SME terms to the quark sector, relevant for Kaon physics, one can bound differences $\Delta a^\mu = a^\mu_{q_1} - a^\mu_{q_2}$, where $q_i, \, i=,2$ 
denote appropriate quark states. The current experimental limits for the coefficients $\Delta_a^\mu$ are: 
from the KTeV Collaboration $\Delta_X, \Delta a _Z <  9.2 \times 10^{-22}$~GeV, while from the 
the KLOE Collaboration in the Da$\Phi$NE $\Phi$ factory~\cite{adidomenico} are less competitive but with the advantage that entangled meson factories have sensitivity to all four coefficients $\Delta a^\mu$, in particular: $\Delta a^0 = (0.4 \pm 1.8 ) \times 10^{-17}$~GeV, from KLOE, with expected sensitivity at KLOE-2 in upgraded DA$\Phi$NE facilties for $\Delta a^{X,Y,Z} = {\mathcal O}(10^{-18})$~GeV. 
Unfortunately, entangled meson factories have only sensitivity to differences $\Delta a^\mu$ rather than absolute coefficients $a^\mu$. Of course, if gravity acts universally for all quark species, such differences may be zero.

\section{Conclusions}
\label{sec:concl}
In this work I have motivated microscopically the existence of some of the SME coefficients that may violate CPT and Lorentz symmetry. In particular, I argued that the presence of Lorentz and CPT Violating geometries in the early universe,  rather than quantum gravity, 
may be responsible for the emergence of $b^\mu$-like axial vector SME backgrounds. Such vectors may be even responsible for the observed matter/antimatter asymmetry in the Universe. From this perspective, having small remnants of such vector backgrounds today is a not so unrealistic possibility, given that the Universe may have undergone a phase transition at a certain temperature during an early era. Hence, it makes perfect sense to search for or bound such SME coefficients by precision atomic spectroscopy or other methods, such as EDMs, including comparison of the relevant properties of matter with antimatter, especially now that we have available man-made antimatter. 

In this talk I reviewed some of the above tests, with direct interest to this conference, including a brief discussion on direct observations of T violation in entangled particle states, independently of CP properties. 
In addition I discussed a novel phenomenon that may characterise certain quantum gravity models, namely ``intrinsic CPT violation'' as a result of the fact that, due to the associated decoherence of matter propagating in a quantum space-time foam environment, the CPT operator is perturbatively ill-defined: although the anti particle exist, nevertheless the properties of the CPT operator when acting on entangled states of particles lead to modified EPR correlators.
Such modifications imply a set of well-defined observables, which can be measured in current or upcoming facilities, such as $\Phi$ or B-factories. 

The signatures of quantum-gravity induced decoherence in entangled states of mesons are rather unique, and in this sense they constitute ``smoking-gun'' evidence for this type of CPT Violation, if realised in Nature. The other important advantage of such searches is that they are virtually cost free, in the sense that the relevant tests can be performed in facilities that have already been or are to be built for other purposes at no extra cost, apart from minor modifications/adjustments  in the relevant Monte-Carlo programmes to take proper account of these quantum-gravity effects.

\begin{acknowledgement}

I would like to thank the Organisers of LEAP 2013 for the invitation to give this plenary talk and for providing an excellently organised and thought-stimulating event.
This work is supported in part by the London Centre for Terauniverse Studies (LCTS), using funding from the European Research Council via the Advanced Investigator Grant 267352 and by STFC (UK) under the research grant ST/J002798/1.

\end{acknowledgement}


\begin{thebibliography}{}
\bibitem{review} For reviews see:   
%%CITATION = ARXIV:0806.0339;%%
D.~Mattingly,
  ``Modern tests of Lorentz invariance,''
  Living Rev.\ Rel.\  {\bf 8}, 5 (2005).
  %%CITATION = 00222,8,5;%%
N.E.~Mavromatos,
  ``CPT violation and decoherence in quantum gravity,''
  Lect.\ Notes Phys.\  {\bf 669}, 245 (2005).
 %%CITATION = LNPHA,669,245;%%


\bibitem{emsH} 
  J.~Ellis, N.~E.~Mavromatos and S.~Sarkar,
  %``Environmental CPT Violation in an Expanding Universe in String Theory,''
  Phys.\ Lett.\ B {\bf 725}, 407 (2013)
  [arXiv:1304.5433 [gr-qc]].
  %%CITATION = ARXIV:1304.5433;%%
  
 \bibitem{polchinski} J.~Polchinski, \textit{String theory } Vols. 1 \& 2 (Cambridge University
Press, 1998).\
%\href{http://www.slac.stanford.edu/spires/find/hep/www?irn=4634802}{SPIRES entry}
  
  
\bibitem{Debnath:2005wk} 
  U.~Debnath, B.~Mukhopadhyay and N.~Dadhich,
  %``Space-time curvature coupling of spinors in early universe: Neutrino asymmetry and a possible source of baryogenesis,''
  Mod.\ Phys.\ Lett.\ A {\bf 21}, 399 (2006)
  [hep-ph/0510351].
  %%CITATION = HEP-PH/0510351;%%
  %13 citations counted in INSPIRE as of 09 Sep 2013

\bibitem{Mukhopadhyay:2005gb} 
  B.~Mukhopadhyay,
  %``Neutrino asymmetry around black holes: Neutrinos interact with gravity,''
  Mod.\ Phys.\ Lett.\ A {\bf 20}, 2145 (2005)
  [astro-ph/0505460].
  %%CITATION = ASTRO-PH/0505460;%%
  %17 citations counted in INSPIRE as of 09 Sep 2013

\bibitem{Sinha:2007uh} 
  M.~Sinha and B.~Mukhopadhyay,
  %``CPT and lepton number violation in neutrino sector: Modified mass matrix of neutrino coupled to gravity,''
  Phys.\ Rev.\ D {\bf 77}, 025003 (2008)
  [arXiv:0704.2593 [hep-ph]].
  %%CITATION = ARXIV:0704.2593;%%
  %10 citations counted in INSPIRE as of 09 Sep 2013

\bibitem{aben}  I.~Antoniadis, C.~Bachas, J.~R.~Ellis and D.~V.~Nanopoulos,
  %``An Expanding Universe in String Theory,''
  Nucl.\ Phys.\ B {\bf 328}, 117 (1989);
  %%CITATION = NUPHA,B328,117;%%
  %258 citations counted in INSPIRE as of 09 Sep 2013
%``Cosmological String Theories and Discrete Inflation,''
  Phys.\ Lett.\ B {\bf 211}, 393 (1988).
  %%CITATION = PHLTA,B211,393;%%
  %263 citations counted in INSPIRE as of 09 Sep 2013

\bibitem{kostel} D.~Colladay and V.~A.~Kostelecky,
  %``Lorentz violating extension of the standard model,''
  Phys.\ Rev.\ D {\bf 58}, 116002 (1998)
  [hep-ph/9809521];
  %%CITATION = HEP-PH/9809521;%%
  %1060 citations counted in INSPIRE as of 10 Sep 2013
V.~A.~Kostelecky,
  %``Lorentz violating and CPT violating extension of the standard model,''
  hep-ph/9912528.
  %%CITATION = HEP-PH/9912528;%%
  %13 citations counted in INSPIRE as of 10 Sep 2013

\bibitem{wheeler} J.~A.~Wheeler and K.~Ford,
  ``Geons, black holes, and quantum foam: A life in physics,''
%\href{http://www.slac.stanford.edu/spires/find/hep/www?irn=4030974}{SPIRES entry}
{\it  New York, USA: Norton (1998)}

\bibitem{lehnert} See:  V.~A.~(.~Kostelecky,``CPT and Lorentz symmetry. Proceedings: 4th Meeting, Bloomington, USA, Aug
8-11, 2007,'') and references therein;
R. Bluhm, V. A. Kostelecky and N. E. Russell, Report No. IUHET-395, AIP Conf. Proc. 457, 70-79 (1999), arXiv:hep-ph/9810327;
S. Coleman and S. L. Glashow, Phys. Rev. D\ \textbf{59}, 116008 (1999), arXiv:hep-ph/9812418.


\bibitem{carrol}  S.~M.~Carroll, J.~A.~Harvey, V.~A.~Kostelecky, C.~D.~Lane and T.~Okamoto,
%  ``Noncommutative field theory and Lorentz violation,''
  Phys.\ Rev.\ Lett.\  {\bf 87}, 141601 (2001).
  %%CITATION = PRLTA,87,141601;%%


\bibitem{greenberg} O.~W.~Greenberg,
%  ``Why is CPT fundamental?,''
  Found.\ Phys.\  {\bf 36}, 1535 (2006).
  %%CITATION = FNDPA,36,1535;%%

\bibitem{chaichian} M.~Chaichian, A.~D.~Dolgov, V.~A.~Novikov and A.~Tureanu,
  %``CPT Violation Does Not Lead to Violation of Lorentz Invariance and Vice Versa,''
  Phys.\ Lett.\ B {\bf 699}, 177 (2011)
  [arXiv:1103.0168 [hep-th]];
  %%CITATION = ARXIV:1103.0168;%%
  %16 citations counted in INSPIRE as of 02 Oct 2013
 M.~Chaichian, K.~Fujikawa and A.~Tureanu,
  %``Lorentz invariant CPT violation,''
  Eur.\ Phys.\ J.\ C {\bf 73}, 2349 (2013)
  [arXiv:1205.0152 [hep-th]].
  %%CITATION = ARXIV:1205.0152;%%
  %3 citations counted in INSPIRE as of 02 Oct 2013

\bibitem{wald} R.~M.~Wald,
%  ``Quantum Gravity And Time Reversibility,''
  Phys.\ Rev.\  D {\bf 21}, 2742 (1980).
  %%CITATION = PHRVA,D21,2742;%%

\bibitem{sarkar} J.~Bernabeu, N.~E.~Mavromatos and S.~Sarkar,
%  ``Decoherence induced CPT violation and entangled neutral mesons,''
  Phys.\ Rev.\  D {\bf 74}, 045014 (2006).
  %%CITATION = PHRVA,D74,045014;%%

\bibitem{millburn} G.~J.~Milburn,
%  ``Lorentz invariant intrinsic decoherence,''
  New J.\ Phys.\  {\bf 8}, 96 (2006).
  %%CITATION = NJOPF,8,96;%%

\bibitem{bmp} J.~Bernabeu, N.~E.~Mavromatos and J.~Papavassiliou,
%  ``Novel type of CPT violation for correlated EPR states,''
  Phys.\ Rev.\ Lett.\  {\bf 92}, 131601 (2004);
  %%CITATION = PRLTA,92,131601;%%
J.~Bernabeu, N.~E.~Mavromatos, J.~Papavassiliou and A.~Waldron-Lauda,
%  ``Intrinsic CPT violation and decoherence for entangled neutral mesons,''
  Nucl.\ Phys.\  B {\bf 744}, 180 (2006).
  %%CITATION = NUPHA,B744,180;%%

\bibitem{smeanti} N. Russell, hep-ph/0209251;
M. Hori and J. Walz, arXiv:1304.3721 and references therein. 

\bibitem{lvqed} See, for instance:  
  P.~A.~Bolokhov and M.~Pospelov,
  %``Classification of dimension 5 Lorentz violating interactions in the standard model,''
  Phys.\ Rev.\ D {\bf 77}, 025022 (2008)
  [hep-ph/0703291 [HEP-PH]], 
  %%CITATION = HEP-PH/0703291;%%
  %46 citations counted in INSPIRE as of 17 Sep 2013
and references therein. 

\bibitem{romalis} P.~A.~Bolokhov, M.~Pospelov and M.~Romalis,
  %``Electric Dipole Moments as Probes of CPT Invariance,''
  Phys.\ Rev.\ D {\bf 78}, 057702 (2008)
  [hep-ph/0609153] and references therein.
  %%CITATION = HEP-PH/0609153;%%
  %3 citations counted in INSPIRE as of 17 Sep 2013


\bibitem{edmfoldy} O.~G.~Kharlanov and V.~C.~.Zhukovsky,
  %``CPT and Lorentz violation effects in hydrogen-like atoms,''
  J.\ Math.\ Phys.\  {\bf 48}, 092302 (2007)
  [arXiv:0705.3306 [hep-th]].
  %%CITATION = ARXIV:0705.3306;%%
  %32 citations counted in INSPIRE as of 26 Sep 2013

\bibitem{dolgov}  See, \emph{e.g}.: A.D.~Dolgov,  Phys. Atom. Nucl., \textbf{73}, 588 (2010), arXiv: 0903.4318, and references therein. 

\bibitem{asacusa} M. Hori \emph{et al.}, Nature \textbf{475}, 484 (2011); doi:10.1038/nature10260. 


\bibitem{adidomenico} M.~Testa  [KLOE Collaboration],
  ``Recent results from KLOE,''
  arXiv:0805.1969 [hep-ex];
  %%CITATION = ARXIV:0805.1969;%%
  F.~Ambrosino {\it et al.}  [KLOE Collaboration],
%  ``First observation of quantum interference in the process Phi $\to$ K(S)  K(L)
%  $\to$ pi+ pi- pi+ pi-: A test of quantum mechanics and CPT symmetry,''
  Phys.\ Lett.\  B {\bf 642}, 315 (2006).
  %%CITATION = PHLTA,B642,315;%%

\bibitem{babar} J.~P.~Lees {\it et al.}  [BaBar Collaboration],
  %``Observation of Time Reversal Violation in the $B^0$ Meson System,''
  Phys.\ Rev.\ Lett.\  {\bf 109}, 211801 (2012)
  [arXiv:1207.5832 [hep-ex]].
  %%CITATION = ARXIV:1207.5832;%%
  %21 citations counted in INSPIRE as of 26 Sep 2013


\bibitem{entangledt} J.~Bernabeu, M.~C.~Banuls and F.~Martinez-Vidal,
  %``$T$ and CPT in $B$ factories,''
  PoS HEP {\bf 2001}, 057 (2001)
  [hep-ph/0111073];
  %%CITATION = HEP-PH/0111073;%%
  %4 citations counted in INSPIRE as of 26 Sep 2013
J.~Bernabeu, F.~Martinez-Vidal and P.~Villanueva-Perez,
  %``Time Reversal Violation from the entangled B0-antiB0 system,''
  JHEP {\bf 1208}, 064 (2012)
  [arXiv:1203.0171 [hep-ph]].
  %%CITATION = ARXIV:1203.0171;%%
  %12 citations counted in INSPIRE as of 26 Sep 2013

\bibitem{kaont} J.~Bernabeu, A.~Di Domenico and P.~Villanueva-Perez,
  %``Direct test of time-reversal symmetry in the entangled neutral kaon system at a \phi-factory,''
  Nucl.\ Phys.\ B {\bf 868}, 102 (2013)
  [arXiv:1208.0773 [hep-ph]].
  %%CITATION = ARXIV:1208.0773;%%
  %2 citations counted in INSPIRE as of 26 Sep 2013
  
  
 



  \bibitem{ehns} J.~R.~Ellis, J.~S.~Hagelin, D.~V.~Nanopoulos and M.~Srednicki,
%  ``Search For Violations Of Quantum Mechanics,''
  Nucl.\ Phys.\  B {\bf 241}, 381 (1984);
  %%CITATION = NUPHA,B241,381;%%
  J.~R.~Ellis, J.~L.~Lopez, N.~E.~Mavromatos and D.~V.~Nanopoulos,
%  ``Precision tests of CPT symmetry and quantum mechanics in the neutral kaon system,''
  Phys.\ Rev.\  D {\bf 53}, 3846 (1996).
  %%CITATION = PHRVA,D53,3846;%%
  P.~Huet and M.~E.~Peskin,
%  ``Violation of CPT and quantum mechanics in the K0 - anti-K0 system,''
  Nucl.\ Phys.\  B {\bf 434}, 3 (1995).
  %%CITATION = NUPHA,B434,3;%%
  For entalged states, the requirement of complete positivity implies a different parametrization for the foam effects (in some cases one may consider $\alpha=\gamma, \beta =0$ in the parameterization of Ellis \emph{et al}.): F.~Benatti and R.~Floreanini,
%  ``Completely positive dynamical maps and the neutral kaon system,''
  Nucl.\ Phys.\  B {\bf 488}, 335 (1997).
  %%CITATION = NUPHA,B488,335;%%
  The experiment can independently measure all three decoherence parameters $\alpha,\beta, \gamma$ of Ellis \emph{et al}. and hence test the assumption of complete positivity, which notably may not be a property of quantum gravity.



\bibitem{bfactories} E.~Alvarez, J.~Bernabeu, N.~E.~Mavromatos, M.~Nebot and J.~Papavassiliou,
%  ``CPT violation in entangled B0 - anti-B0 states and the demise of  flavour
%  tagging,''
  Phys.\ Lett.\  B {\bf 607}, 197 (2005);
  %%CITATION = PHLTA,B607,197;%%
E.~Alvarez, J.~Bernabeu and M.~Nebot,
%  ``Delta(t)-dependent equal-sign dilepton asymmetry and CPTV effects in  the
%  symmetry of the B0 anti-B0 entangled state,''
  JHEP {\bf 0611}, 087 (2006).
  %%CITATION = JHEPA,0611,087;%%

\end{thebibliography}
\end{document}